\documentclass[hidelinks]{styles/svproc}
\usepackage{url}

\usepackage{amsmath,amssymb,amsfonts}
\usepackage{orcidlink}
\usepackage{flafter} 
\usepackage{placeins} 
\usepackage{hyperref} 
\usepackage{subcaption}
\usepackage{makecell}
\usepackage{multirow}

\definecolor{UStuttDarkBlue}{RGB}{0,81,158}
\definecolor{UStuttDarkOrange}{RGB}{228,175,52}

\usepackage{tikz}
\usetikzlibrary{calc,fit, positioning,arrows.meta}
\tikzset{>={Latex[width=2mm,length=2mm]}} 
\tikzstyle{block} = [draw=black, fill=white, rectangle, align=center, minimum height=1em, minimum width=2em, font = {\footnotesize}]
\tikzstyle{sum} = [draw, circle, node distance=1cm]
\tikzstyle{input} = [coordinate]
\tikzstyle{output} = [coordinate]
\tikzstyle{punkt} = [circle,fill,inner sep=0pt,minimum size=5pt]
\tikzstyle{gray dotted} = [draw=gray, fill=white, rectangle, align=center, line width=1pt, dash pattern=on 1pt off 4pt on 6pt off 4pt, inner sep=3mm, rounded corners, font = {\small}]
\tikzstyle{blueblock} = [draw=UStuttDarkBlue, fill=white, rectangle, align=center, minimum height=1em, minimum width=2em, line width =1.5pt, font = {\footnotesize}]
\tikzstyle{orangeblock} = [draw=UStuttDarkOrange, fill=white, rectangle, align=center, minimum height=1em, minimum width=2em, line width =2pt, font = {\footnotesize}]
\newcommand{\markcircle}[3]{
	\node[
	circle,
	fill={rgb,255:red, 0; green, 190; blue, 255},
	text=white,
	minimum size=4mm,
	] at (#1,#2) {#3};
}

\usepackage{siunitx} 
\DeclareSIUnit{\Bit}{Bit}

\begin{document}
\mainmatter              
\title{Transmission Error Compensation for Industrial Robots Using Fourier Series}
\titlerunning{Transmission Error Compensation}  
%
\author{Christian~J.\,E.~Bauer\inst{1}~\orcidlink{0009-0002-7979-3877}\and
	Simon~Nowinski\inst{1}~\orcidlink{0009-0000-9545-5371}\and
	Valentin~Kamm\inst{1}~\orcidlink{0000-0003-4748-5415}\and
	Lukas~Steinle\inst{1}~\orcidlink{0000-0002-6035-0067}\and
	Armin~Lechler\inst{1}~\orcidlink{0000-0002-4073-1487}\and
	Alexander~Verl\inst{1}~\orcidlink{0000-0002-2548-6620}}
\authorrunning{Christian J.\,E. Bauer et al.} 
%
\tocauthor{Christian J. E. Bauer, Simon Nowinski, Valentin Kamm, Lukas Steinle, Armin Lechler, Alexander Verl}

\institute{Institute for Control Engineering of Machine Tools and Manufacturing Units~(ISW),\\ University of Stuttgart, 70174 Stuttgart, Germany,\\
\email{christian.bauer@isw.uni-stuttgart.de},\\
\texttt{http://isw.uni-stuttgart.de}}

\maketitle              
\begin{abstract}
Due to their low accuracy, industrial robots commonly require online teaching after the offline development of a robot program in simulation.
The accuracy is limited by compliance and transmission errors in the robot joints.
Transmission errors are defined as errors between the load-side angles of the gearboxes and the motor encoder angles.
An approach to identify the transmission errors based on laser tracker measurements is presented.
The developed velocity-dependent Fourier series model is integrated in a virtual commissioning software as a co-simulated functional mock-up unit.
Additionally, it is used to compensate the transmission errors by correcting the setpoint from the computerized numerical control (CNC).
The model identification and the compensation are experimentally evaluated with a Stäubli TX2-40 industrial robot.
The model has a mean absolute error of 0.114 mrad, and the compensation improves the accuracy by 83.26 \% for axis 2 and 49.21 \% for axis 3.
\end{abstract}

\section{Introduction}
The programming of industrial robots~(IR) currently heavily relies on online teaching, where only repeatability is relevant.
Although the usage of digital twins for offline programming is common, the online adaptation of previously offline prepared robot programs is still necessary due to insufficient accuracy of standard IRs.
For example, a mean error of \qty{0.968}{\mm} and a maximum error of \qty{2.158}{\mm} were reported for an ABB~IRB~1600 with a reach of~\qty{1.45}{\m}~\cite{nubiolaAbsoluteCalibrationABB2013}, while a mean error of \qty{4.257}{\mm} and a maximum error of~\qty{5.751}{\mm} were measured for a KUKA~KR210-2 with a reach of~\qty{2.7}{\m}~\cite{dzubbaNonlinearElasticityModel2025}.
Since the models for the inaccuracy of standard IRs are usually not available, these inaccuracies are neither considered in digital twins, nor compensated in the robot control.
This increases overall commissioning time, and therefore, commissioning cost.
Due to this limited accuracy and the restricted offline programming capabilities, IRs are mainly used for handling applications, which do not require frequent changes in the robot program.
In contrast, manufacturing tasks require high accuracy and the ability for flexible reprogramming.
Therefore, robot models with high accuracy are needed to enable the direct use of offline developed robot programs.
Furthermore, they are required to increase the accuracy of an IR to a suitable level for manufacturing processes with low to medium accuracy demands.
The path errors of IRs mainly originate from the robot joints consisting of permanent-magnet synchronous motors~(PMSM) and gearboxes~\cite{olabiImprovingAccuracyIndustrial2012}.
Within these drive trains, the angular position error is caused by gearbox compliance, kinematic transmission errors~(TEs), and motor torque ripples~\cite{rendersKinematicCalibrationGeometrical1991}.
The errors of individual joints are accumulated along the serial kinematic chain and amplified by the length of the arm to the tool center point~(TCP).
TEs cannot be compensated for by the position controller because encoders in standard IRs are mounted at the motor shaft.
A TE model for cycloidal drives considering load-, velocity-, and temperature-dependency was previously presented by the authors~\cite{bauerModelingLoadVelocity2025}.
This paper aims at presenting a first approach to identify TEs for standard IRs using a laser tracker, transferring the modeling method, and showing the potential of improving digital twins by more accurate, dynamic robot joint models.
Deviations of the kinematic robot description are not considered here.

\subsection{State of the Art}
\label{ssec:StateOfTheArt}
TEs are defined as errors between motor and load-side angles of the gearbox.
These errors are angle-dependent and periodic~\cite{blancheCycloidDrivesMachining1989}.
TEs are caused by manufacturing and assembly tolerances.
In previous works, TEs were mainly modeled using angle-dependent Fourier series.
\textsc{Ghorbel} et~al. additionally considered the load-torque and velocity dependency~\cite{ghorbelKinematicErrorHarmonic2001}.
Furthermore, a lubricant temperature dependency was described, which can be explained by changes in the viscous friction~\cite{qiuReviewPerformanceTesting2021}.
In literature, many works cover the modeling of TEs in strain wave gears~\cite{iwasakiModelingCompensationAngular2009,hanSuppressionVibrationDue2008,iwasakiVibrationSuppressionAngular2014}, while there is less research covering TEs in cycloidal drives~(CDs)~\cite{zhangMathematicalModelAnalysis2011}.
Disturbance observers for strain wave gears~\cite{iwasakiModelingCompensationAngular2009} and for CDs~\cite{kawaharaVibrationSuppressionFeedback2013} based on purely angle-dependent Fourier series have been shown.
\textsc{Bilancia} et~al. reported the velocity and temperature dependency of TEs also for CDs~\cite{bilanciaAccurateTransmissionPerformance2022}.
In~\cite{bilanciaOnlineMotionAccuracy2025}, \textsc{Bilancia} et~al. modeled and compensated the influence of load-, velocity-, and temperature-dependency using machine learning models with the load-side angle as an input.
\textsc{Bauer} et~al. use a load-, velocity-, and temperature-dependent Fourier series with the motor angle as an input to model TEs in CDs~\cite{bauerModelingLoadVelocity2025}.
\textsc{Han} et~al. applied an acceleration sensor with a peak filter to suppress vibrations due to TEs in an industrial robot~\cite{hanSuppressionVibrationDue2008}, while \textsc{Iwasaki} et~al. corrected the torque setpoint with an adaptive notch filter~\cite{iwasakiVibrationSuppressionAngular2014}.
Moreover, there are several approaches based on load-side encoders, incorporating H-infinity control, unscented Kalman filters~\cite{mesmerDriveBasedVibrationDamping2020}, and gain scheduling~\cite{mesmerGainScheduledDrivebasedDamping2022}.
\textsc{Olabi} et~al. analyzed the TEs of a Stäubli~RX~170~BH based on laser tracker measurements of the load-side angles.
The authors identified the stiffness of the joints by attaching constant weights and measuring the deformation.
The TEs of axis~\num{2} were compensated by adapting the motor angle setpoint using a lookup table~\cite{olabiImprovingAccuracyIndustrial2012}.
\textsc{Vakilinejad} et~al. analyzed the TEs of the first three joints of a Stäubli~TX90 under the assumption that the first three joints contribute most to the TCP error due to their larger lever arm and higher gear ratios.
The authors model the influence of gravity load and runout error as third-order polynomial and the periodic TEs as angle-dependent sinus series with seven harmonics~\cite{vakilinejadIdentificationCompensationPeriodic2019}. 

\subsection{Contribution}
First, this paper aims at presenting a method to identify the velocity-dependent TEs from laser tracker and motor encoder measurements.
The laser tracker measurements are captured in the Cartesian laser tracker coordinate frame and have to be transformed to load-side angles.
A Stäubli~TX2-40~(\autoref{fig:Staeubli}~(a)) with a reach of~\qty{0.515}{\m} and a load capacity of~\qty{2}{\kg} is used for experiments~\cite{TX2406axisProductdatasheetEN}.
Second, the integration of accurate TE models in the digital twin of robot cells is demonstrated.
ISG-virtuos is used as deterministic real-time simulation tool. Here, the TE models are integrated as co-simulated functional mock-up units~(FMUs) with a non-real-time solver.
During the virtual commissioning, the simulated robot considering the TEs is shown as a shadow model over the purely kinematic simulation, highlighting the differences in modeling complexity~(\autoref{fig:Staeubli}~(b)).
Third, the derived models are integrated as feedforward control to compensate the TEs. This proposes an alternative to the consideration of TEs during offline programming.
The compensation is implemented as a feedforward control in TwinCAT~3.
It corrects the angle setpoint from the computerized numerical control~(CNC) in the programmable logic controller~(PLC) and sends the corrected setpoint via EtherCAT and the uniVAL interface to the robot.
The TEs were only identified for the first three axes of the robot, since it is assumed that they contribute most to the TCP error due to a larger lever arm~\cite{vakilinejadIdentificationCompensationPeriodic2019}.
Furthermore, only TEs including compliance were identified and compensated within this contribution.
To increase the overall accuracy at the TCP, the additional identification of deviations from the kinematic description provided by the manufacturer is necessary.

\section{Experimental Setup and Methods}
\begin{figure}
	\centering
	\begin{subfigure}{0.495\textwidth}
		\centering
		\begin{tikzpicture}
			\node[anchor=south west, inner sep=0] (img)
			{\includegraphics[scale=0.15]{"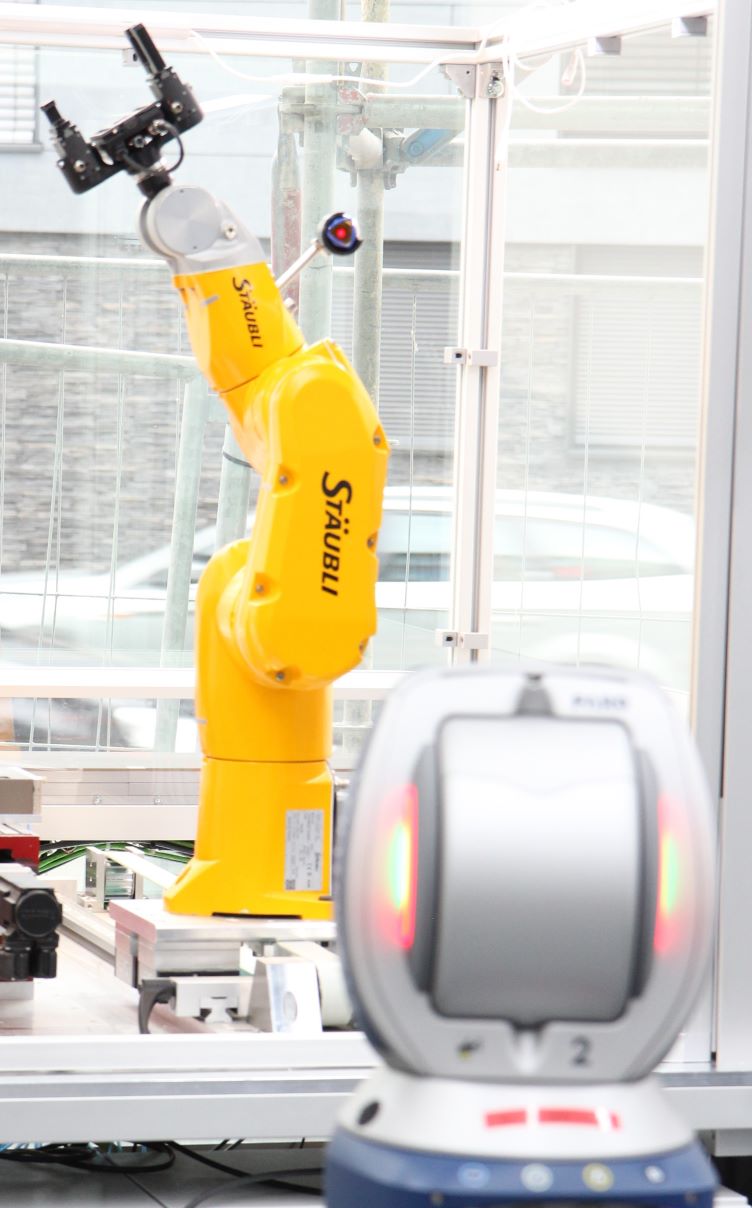"}};
			\begin{scope}[x={(img.south east)}, y={(img.north west)}]
				\markcircle{0.36}{0.3}{1}
				\markcircle{0.72}{0.28}{2}
				\markcircle{0.58}{0.83}{3}
			\end{scope}
		\end{tikzpicture}
		\caption{Stäubli~TX2-40~(1), with FARO laser tracker~(2), and laser tracker target~(3)}
	\end{subfigure}
	\hfill
	\begin{subfigure}{0.495\textwidth}
		\centering
		\includegraphics[scale=0.2]{"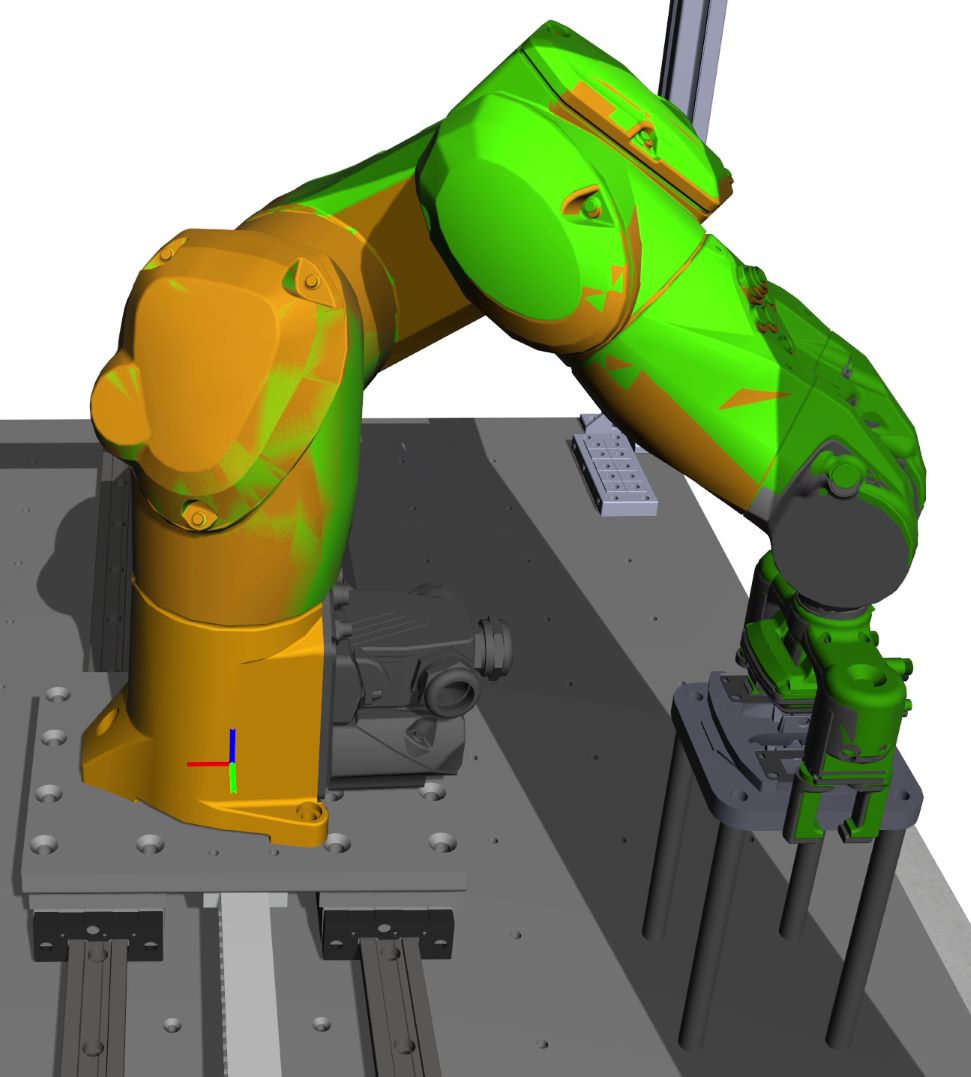"}
		\caption{Shadow model in ISG-virtuos}
	\end{subfigure}
	\caption{Measurement of the load-side angle with a laser tracker, and visualization of position offset due to TEs within the simulation for offline programming.}
	\label{fig:Staeubli}
\end{figure}

The Stäubli TX2-40 used in the experiments is shown in \autoref{fig:Staeubli}~(a).
The load-side angle of the first three joints was derived from measurements with a FARO Vantage~E6 laser tracker.
It recorded the Cartesian position of a MetrologyWorks~\num{1.5}" spherically mounted retroreflector~(SMR) installed above the fourth joint.
\label{ssec:experimental_setup}
The robot has~\num{6} degrees of freedom and a repeatability of~$\pm\qty{0.02}{\mm}$ according to ISO~9283~\cite{TX2406axisProductdatasheetEN}.
In the first two joints are Stäubli-patented Joint Combiné Stäubli~(JCS)~52~M cycloidal drives~\cite{US_5286235_A_misc,US_5312305_A_misc}.
Joint three incorporates a worm gear with the reduction ratio~\num{45}, while the worm gears in joint four to six have a reduction ratio of~\num{35}.
The control system is running at~\qty{1}{\kilo\Hz} cycle time.

\subsection{Modeling of the Transmission Errors}
According to~\cite{bauerModelingLoadVelocity2025}, TEs are defined as~\(\theta_\mathrm{TE} = q - \theta\).
Using this definition, compliance is included in the TEs.
In this contribution, the input-side angle~\(\theta\) is scaled by the reduction ratio to facilitate comparability with the load-side angle~\(q\).
The TEs are modeled by a Fourier series in the amplitude-phase form
\begin{align}
	\label{eq:hat_theta_TE}
	\hat{\theta}_\mathrm{TE} (\theta, \dot{\theta}) &=
	\sum_k {A_k \cdot
		\sin \left( f_k \cdot \theta + \varphi_k \right) }
	+ a_3 \cdot \theta^3 + a_2 \cdot \theta^2 + a_1 \cdot \theta + a_0~\mathrm{,}
\end{align}
including~\(k\) harmonic frequencies~\(f_k\) as well as amplitude~\(A_k\) and phase~\(\varphi_k\).
Here, the spatial frequencies~\(f_k\) are dimensionless multiples based on a full load-side revolution~(period~\(\theta_\mathrm{period} = 2\pi\)).
An error component~\(k\) repeats~\(f_k\)-times per load-side revolution.
Due to obstacles in the workspace and restrictions by the cable drag chain, it is not possible to record TE data for a full load-side revolution, as in~\cite{bauerModelingLoadVelocity2025}.
A data analysis has shown that this prevents a robust identification of a Fourier series model for the low frequency TE components.
Therefore, a third order polynomial with the parameters~\(a_3\),~\(a_2\),~\(a_1\), and~\(a_0\) is used to model the runout error and low frequency TE components~\cite{vakilinejadIdentificationCompensationPeriodic2019}.
For lightweight robots with low gravity load in applications without process forces, it is assumed to be sufficient to consider only the velocity-dependency of TEs.
Consequently,~\(A_k\),~\(\varphi_k\), and~\(a_2\) to~\(a_0\) are approximated as quadratic functions of the form
\begin{align}
	\label{eq:A_k}
	A_k \left(\dot{\theta}\right)&= p_{A,k,2} \cdot  \dot{\theta}^2 + p_{A,k,1} \cdot \dot{\theta} + p_{A,k,0}~\mathrm{,}\\
	\label{eq:phi_k}
	\varphi_k \left(\dot{\theta}\right)&= p_{\varphi,k,2} \cdot \dot{\theta}^2 + p_{\varphi,k,1} \cdot \dot{\theta}+ p_{\varphi,k,0}~\mathrm{and} \\
	\label{eq:a}
	a_{\cdot} \left(\dot{\theta}\right)&= p_{a \cdot,2} \cdot \dot{\theta}^2 + p_{a \cdot,1} \cdot \dot{\theta}+ p_{a \cdot,0}~\mathrm{,}
\end{align}
with the coefficients~\(p\).
To further improve the TE estimation, additional dependencies on load torque~\(\tau_\mathrm{load}\) and temperature~\(T\) can be included~\cite{bauerModelingLoadVelocity2025}.

\subsection{Extraction of Load-Side Angles from Cartesian Measurements}
\label{ssec:extractLoadSide}
To identify the model parameters, each axis was moved along an ideally circular arc with constant joint velocity.
The load-side angle of each joint was calculated from laser tracker measurements recorded as point set \(\{\mathbf{p}_i \in \mathbb{R}^3\}_{i=1}^N\).
The data was centered by subtracting the mean~\(\bar{\mathbf{p}} \in \mathbb{R}^3\). A singular value decomposition~(SVD) of the centered data matrix
\begin{align}
\mathbf{P}_0 &= \mathbf{U}\boldsymbol{\Sigma}\mathbf{V}^\top~\mathrm{,}
\end{align}
is computed.
The right singular vector corresponding to the smallest singular value defines the unit normal \(\mathbf{n} \in \mathbb{R}^3\) of the best-fit plane in the total least-squares sense.
The points are orthogonally projected onto this plane to receive the projected points~$\tilde{\mathbf{p}}_i \in \mathbb{R}^3$.
Using the first two right singular vectors as an orthonormal in-plane basis, the projected points are expressed in 2D coordinates~\(\mathbf{u}_i \in \mathbb{R}^2\).
A circle fitting~\(\mathbf{u}_i\) is estimated, applying the linear least squares method, and 2D circle center~\(\mathbf{c}_2 \in \mathbb{R}^2\) and radius~\(r\) are obtained.
The 3D center~$\mathbf{c}_3 \in \mathbb{R}^3$ is received by back-transformation into the original coordinate system.
After estimating circle center~\(\mathbf{c}_3\) and plane normal~$\mathbf{n}$,
each projected point~$\tilde{\mathbf{p}}_i$ is expressed relative to the center via the normalized radial vector
\begin{align}
	\mathbf{v}_i &= \frac{\tilde{\mathbf{p}}_i - \mathbf{c}_3}{\|\tilde{\mathbf{p}}_i - \mathbf{c}_3\|}~\mathrm{.}
\end{align}
The first point of an individual measurement defines a normalized reference direction~\(\mathbf{v}_1\).
For each point~\(i\), the load-side angle~\(q_i \in (-\pi, \pi]\) is computed by
\begin{align}
	q_i &= \mathrm{atan2}\left(
	\mathbf{n}^\top \cdot (\mathbf{v}_1 \times \mathbf{v}_i),
	\mathbf{v}_1^\top \cdot \mathbf{v}_i
	\right)~\mathrm{.}
\end{align}

\subsection{Parameter Identification}
\label{ssec:identification}
During the identification measurements, the joint angle setpoint~\(q_\mathrm{set}\) was commanded to one axis at a time using~\num{13} discrete, symmetric velocity levels:
\begin{align*}
	q_\mathrm{set,1} &\in [-145^\circ, 170^\circ]~\mathrm{,} \quad
	q_\mathrm{set,2} \in [-100^\circ, 95^\circ]~\mathrm{,} \quad
	q_\mathrm{set,3} \in [-80^\circ, 135^\circ]~\mathrm{,} \\
	\dot{q}_{\mathrm{set},m} &\in \{\pm(1.67 + 3.33s), \quad s = 0,1,\dots,\qty{12}{\degree\per\s}\},
	\quad m = \{1,2,3\}~\mathrm{.}
\end{align*}
The actual load-side angle~\(q\) was derived from the laser tracker measurements, applying the method from~\autoref{ssec:extractLoadSide}, while the encoder angles~\(\theta\) were directly recorded in TwinCAT.
In the data processing, time offset between the measurement data was estimated by matching the rising and falling edge of the velocity.
For each measurement at each velocity, the polynomial parameters~\(a_2\),~\(a_1\), and~\(a_0\) were estimated using the linear least squares method.
The resulting third-order polynomial was subtracted from the TE data~\(\theta_\mathrm{TE}\), and~\(A_k\) and~\(\varphi_k\) in \autoref{eq:hat_theta_TE} are calculated for each chosen~\(f_k\) as described in~\cite{bauerModelingLoadVelocity2025}.
In the final processing step, the model parameters~\(p\) were calculated by applying the linear least squares method to solve \autoref{eq:A_k} to \autoref{eq:a}.

\subsection{Compensation Approach}
The developed compensation needs the angle setpoint from the CNC as input~(\autoref{fig:BlockDiagramOverview}).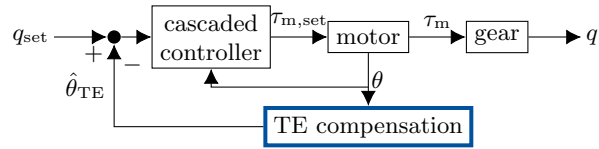
\begin{figure}[b]
	\centering
	\begin{tikzpicture}
	\node[punkt] (sum1) {};
	\node[block, right=0.4cm of sum1] (controller) {cascaded\\controller};
	\node[block, right=0.75cm of controller] (motor) {motor};
	\node[block, right=0.75cm of motor] (gear) {gear};
	
	\node[blueblock, below=0.7cm of motor] (compensation) {TE compensation};
	
	\node[input, left=0.7cm of sum1] (input1) {};
	\node[input, right=0.7cm of gear] (output1) {};
	
	\draw[->] (input1) -- node[near end, below] {\(+\)} node[left, xshift=-0.3cm] {\(q_\mathrm{set}\)} (sum1);
	\draw[->] (sum1) -- (controller);
	\draw[->] (controller) -- node[midway, above] {\(\tau_{\mathrm{m,set}}\)} (motor);
	\draw[->] (motor) -- node[midway, above] {\(\tau_{\mathrm{m}}\)} (gear);
	\draw[->] (gear) -- node[right, xshift=0.3cm] {\(q\)} (output1);
	\draw[->] (motor) -- node[midway, right, xshift=-0.075cm] {\(\theta\)} (compensation);
	\draw[->] (compensation) -| node[near end, left] {\(\hat{\theta}_\mathrm{TE}\)} node[very near end, right] {\(-\)} (sum1);
	
	\draw[->] ($(motor.south)+(0,-0.45)$) -| (controller);
\end{tikzpicture}
	\caption{Compensation approach with motor torque~\(\tau_{\mathrm{m}}\).}
	\label{fig:BlockDiagramOverview}
\end{figure}
The required velocity setpoint is not derived from the CNC, but instead calculated internally.
To avoid abrupt changes in the compensation output, an exponential moving average filter with a forgetting factor of~\num{0.99} is applied to that velocity.
The derived TE compensation offset is subtracted from the angle setpoint and commanded to the Stäubli robot.
During the activation of the compensation, a seven phase profile is used to ensure a smooth transition over~\qty{5}{s}.
As reference for evaluation, the theoretical motor angle without compensation is calculated as angle setpoint minus controller error.

\section{Results}
\label{sec:results}
\begin{figure}[tb]
	\centering
	\begin{subfigure}{0.495\textwidth}
		\includegraphics[scale=1]{"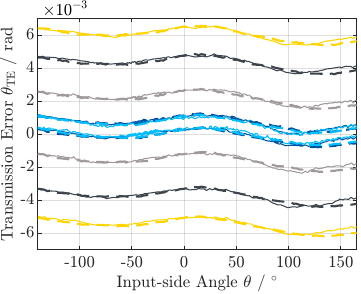"}
		\caption{Position domain}
	\end{subfigure}
	\hfill
	\begin{subfigure}{0.495\textwidth}
		\includegraphics[scale=1]{"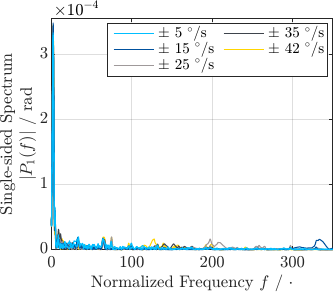"}
		\caption{Spatial frequency domain}
	\end{subfigure}
	\caption{TE measurements of axis 1 and the model estimation (dashed lines).}
	\label{fig:axis_1}
\end{figure}

\begin{figure}
\centering
\begin{subfigure}{0.495\textwidth}
	\includegraphics[scale=1]{"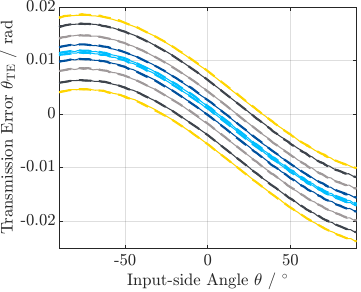"}
	\caption{Position domain}
\end{subfigure}
\hfill
\begin{subfigure}{0.495\textwidth}
	\includegraphics[scale=1]{"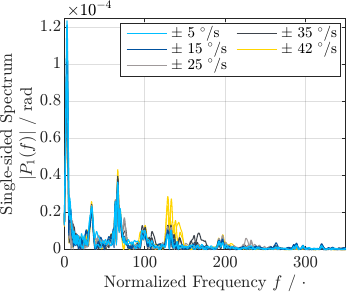"}
	\caption{Spatial frequency domain}
\end{subfigure}
\caption{TE measurements of axis 2 and the model estimation (dashed lines).}
\label{fig:axis_2}
\end{figure}

\begin{figure}
\centering
\begin{subfigure}{0.495\textwidth}
	\includegraphics[scale=1]{"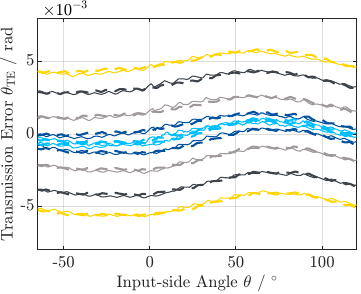"}
	\caption{Position domain}
\end{subfigure}
\hfill
\begin{subfigure}{0.495\textwidth}
	\centering
	\begin{tikzpicture}
		\node[anchor=south west, inner sep=0] (img) at (0,0)
		{\includegraphics[width=\linewidth]{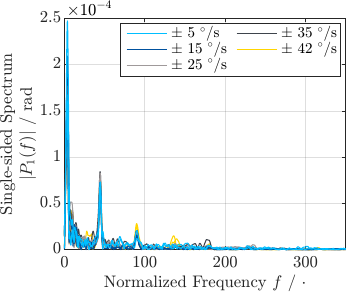}};
		
		\begin{scope}[x={(img.south east)}, y={(img.north west)}]
			
			\draw[->, black, thick] (0.47,0.63) -- (0.2875,0.42)
			node[very near start, above=0.1cm] {Reduction ratio 45};
			
			\coordinate (A) at (0.7,0.45);
			
			\draw[->, black, thick] (A) -- (0.395,0.245)
			node[very near start, above=0.125cm, text width=3cm, align=center, xshift=0.3cm] {Harmonics of reduction ratio};
			\draw[->, black, thick] (A) -- (0.4975,0.2025);
			\draw[->, black, thick] (A) -- (0.6,0.1875);
		\end{scope}
	\end{tikzpicture}
	\caption{Spatial frequency domain}
\end{subfigure}
\caption{TE measurements of axis 3 and the model estimation (dashed lines).}
\label{fig:axis_3}
\end{figure}

\begin{figure}
	\centering
		\includegraphics[scale=1]{"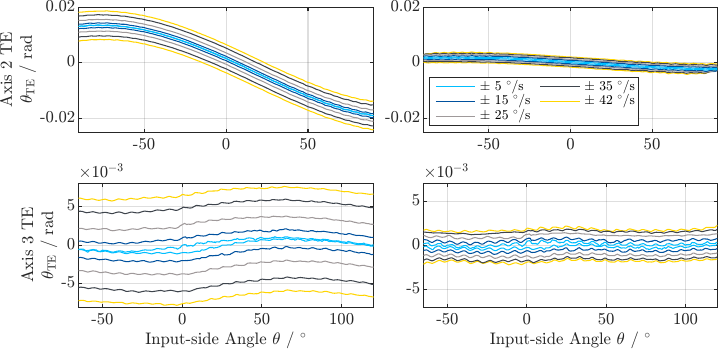"}
	\begin{subfigure}{0.495\textwidth}
		\caption{Without compensation}
	\end{subfigure}
	\hfill
	\begin{subfigure}{0.495\textwidth}
		\caption{With compensation}
	\end{subfigure}
	\caption{TE of axis 2 and 3 without and with compensation.}
	\label{fig:compensation}
\end{figure}

The measured transmission error and its estimation by the model for the first three axes are shown in \autoref{fig:axis_1} to \autoref{fig:axis_3} for selected measurements.
The identification data and figures for all measurements can be found in the dataset~\cite{DARUS-5760_2026}.
The analysis in the spatial frequency domain shows a dominant peak at low frequencies, which indicates that stiffness and runout error contribute most to the overall~\(\theta_\mathrm{TE}\).
Axis~\num{1} and~\num{2} show the most dominant peaks approximately at the normalized spatial frequency~\num{33} and its harmonics.
Therefore, analogous to the results in~\cite{bauerModelingLoadVelocity2025}, it is assumed that the reduction ratio of these gearboxes is~\num{33}.
For axis~\num{3}, the dominant peaks in the spatial frequency domain are at~\num{45} and its first four harmonics, which aligns with the known reduction ratio of~\num{45}.
The most dominant frequencies were selected for model identification~(as in~\cite{bauerModelingLoadVelocity2025}), yielding \num{14}, \num{16}, and \num{8} frequencies for axes~\num{1}–\num{3}, respectively.
The model output shows the root mean square error~(RMSE), mean absolute error~(MAE) and a maximum error listed in \autoref{tab:errors} with respect to the measurement data.\begin{table}
\caption{Errors of Model, TE Without and With Compensation}
\label{tab:errors}
\centering
\begin{tabular}{l c c c c c c}
	\hline
	\rule{0pt}{12pt}
	& Axis & 
	\makecell{RMSE\\(\si{\radian})} & 
	\makecell{MAE\\(\si{\radian})} & 
	\makecell{Maximum\\(\si{\radian})} & 
	\makecell{MAE TCP\\(\si{mm})} & 
	\makecell{Maximum\\TCP (\si{mm})} \\
	\hline
	
	\rule{0pt}{12pt}
	\multirow{3}{*}{Model error} 
	& \num{1} & \num{1.52e-4} & \num{1.22e-4} & \num{0.78e-3} & \num{6.29e-2} & \num{0.40} \\
	& \num{2} & \num{0.88e-4} & \num{0.69e-4} & \num{0.81e-3} & \num{3.55e-2} & \num{0.42} \\
	& \num{3} & \num{1.80e-4} & \num{1.52e-4} & \num{1.00e-3} & \num{4.41e-2} & \num{0.29} \\
	\hline
	
	\rule{0pt}{12pt}
	\multirow{2}{*}{\makecell{TE without\\compensation}} 
	& \num{2} & \num{10.54e-3} & \num{9.20e-3} & \num{23.68e-3} & \num{4.74} & \num{12.20} \\
	& \num{3} & \num{1.99e-3} & \num{1.91e-3} & \num{5.83e-3} & \num{0.55} & \num{1.69} \\
	\hline
	
	\rule{0pt}{12pt}
	\multirow{2}{*}{\makecell{TE with\\compensation}} 
	& \num{2} & \num{1.82e-3} & \num{1.54e-3} & \num{4.25e-3} & \num{0.79} & \num{2.19} \\
	& \num{3} & \num{1.00e-3} & \num{0.97e-3} & \num{5.39e-3} & \num{0.28} & \num{1.56} \\
	\hline
\end{tabular}
\end{table}
In \autoref{fig:compensation}, the resulting TEs after applying the proposed compensation approach in the control is shown.
For axis~\num{2}, the accuracy is improved by~\qty{83.26}{\percent} with respect to the MAE.
Axis~\num{3} shows a relative improvement of~\qty{49.21}{\percent}.
The MAE and the maximum error at the TCP are estimated based on the TEs and the maximum distance~\(d\) from joint to TCP~(axis~\num{1} \&~\num{2}:~\qty{515}{\mm}, axis~\num{3}:~\qty{290}{\mm}~\cite{TX2406axisProductdatasheetEN}).
The error at the TCP~\(\Delta_\mathrm{TCP}\) is derived under the simplified assumption of an isosceles triangle~\(\Delta_\mathrm{TCP}=2 d \sin (0.5\theta_\mathrm{TE})\).
These error values indicate how much the TEs of an individual axis contribute to the overall TCP error.
At the TCP, the MAE of axis~\num{2} shows a reduction from~\qty{4.74}{\mm} to~\qty{0.79}{\mm}. For axis~\num{3} the MAE is reduced from~\qty{0.55}{\mm} to~\qty{0.28}{\mm}.

\section{Discussion and Outlook}
The presented approach proved its suitability to identify TEs of robot gearboxes, and its ability to reduce these at least by half.
However, the model output still shows deviations of up to~\qty{1.00e-3}{\radian} from the measurement.
This can be explained by the neglected load- and temperature-dependency of the TEs.
Especially, the load-dependency has a strong influence~\cite{bauerModelingLoadVelocity2025}, which was only implicitly considered here by the current pose.
Since the gravity load of the robot does not change for the same pose, the load during identification and compensation measurement was similar.
In the evaluation experiments, the TEs were not completely compensated, which indicates that the model underestimated the TEs.\\
While the presented approach can improve the accuracy of the individual joints of the robot, it is necessary to identify the deviations of the kinematics from the ideal robot in an additional step, to achieve the overall goal of applying offline-developed robot programs directly.
To allow a robust modeling and compensation of the TEs, the model could be extended by an inverse dynamics model.
Assuming a constant density, a simple approach would be to identify a load model based on the computer-aided design~(CAD) model.
To provide a wider variety of identification data, measurements with different additional payloads are desirable.
As presented in~\cite{bauerModelingLoadVelocity2025}, the Fourier series can be extended by a load- and temperature-dependency by adapting \autoref{eq:A_k} to \autoref{eq:a}.
Furthermore, it would be possible to attach temperature sensors to the housing of the robot gearboxes to provide the temperature input for the TE model~\cite{bauerLubricantTemperatureObserver2025}.

\section*{Data Availability and Acknowledgment}
The data used to identify the TE model is available on DaRUS, the data repository of the University of Stuttgart:
\url{https://doi.org/10.18419/DARUS-5760}\\
This project was funded by the Deutsche Forschungsgemeinschaft (DFG, German Research Foundation, 443677015).

\enlargethispage{3\baselineskip}

%
%
\bibliographystyle{styles/bibtex/spmpsci_unsrt}
\bibliography{2026_MHI, 2026_MHI_manual, ICIT_2025_BibTeX}
\end{document}